\documentclass[11pt,english,aps,tightenlines,prb,showpacs]{revtex4}
\usepackage{times}
\usepackage[T1]{fontenc}
\usepackage[latin1]{inputenc}
\usepackage{amsmath}
\usepackage{babel}
\usepackage{graphics}
\usepackage{amssymb}
\usepackage{subfigure}

\makeatletter

\providecommand{\LyX}{L\kern-.1667em\lower.25em\hbox{Y}\kern-.125emX\@}

\DeclareMathOperator{\erf}{erf}
\DeclareMathOperator{\erfi}{erfi}

\makeatother
\begin{document}


\title{The classical Generalized Constant Coupling method for Geometrically
Frustrated Magnets revisited: Microscopic formulation and effect of
perturbations beyond nearest neighbor interactions.}

\author{Angel J. Garcia-Adeva}

\email{garcia@landau.physics.wisc.edu}

\author{David L. Huber}

\affiliation{Department of Physics; University of Wisconsin--Madison; Madison,
WI 53706}

\begin{abstract}
A microscopic derivation of the classical Generalized Constant Coupling
(GCC) model for geometrically frustrated magnets is presented. Starting
from the classical Heisenberg Hamiltonian, the partition function
for clusters with \( p=2,\, 3,\, 4,\ldots  \) spins in the presence
of the inhomogeneous symmetry breaking fields (SBF) created by spins
outside the unit is calculated. The effective fields characterizing
the interaction between units naturally arise as averages over the
SBF. These effective fields are fixed by a self-consistency condition.
In the paramagnetic regime, we recover all the results previously
obtained in a more phenomenological way, which were shown to be in
excellent agreement with Monte Carlo calculations for these lattices.
However, this microscopic formulation allows us to also study the
behavior of the system in the absence of applied magnetic field. It
is found that, for antiferromagnetic interactions, the equilibrium
configuration is a non-collinear configuration in which the total
magnetization of the unit is zero, and the condition under which such
an ordered state occurs is also obtained from the calculation. However,
frustration inhibits the formation of such an state, and the system
remains paramagnetic down to 0 K, if only nearest neighbor interactions
are taken into account, for all the systems considered, in agreement
with the now generally accepted idea. For completeness, the ferromagnetic
case is also studied. Furthermore, as the present method is formulated
in the real space, it is very easy to study the effect of further
perturbations in the Hamiltonian by modifying the form of the SBF.
We do this to include the effect of next nearest neighbor interactions
(NNN) and site dilution by non-magnetic impurities for the pyrochlore
lattice. It is found that NNN interactions can stabilize a non-collinear
ordered state, or ferromagnetic one, depending on the relation between
NN and NNN interactions, in agreement with mean field calculations,
and the phase diagram is calculated. However, site dilution is not
enough by itself to form such an ordered state.
\end{abstract}
\maketitle

\section{Introduction}

\begin{figure}[htb!]
{\centering \subfigure[Kagome lattice]{\includegraphics{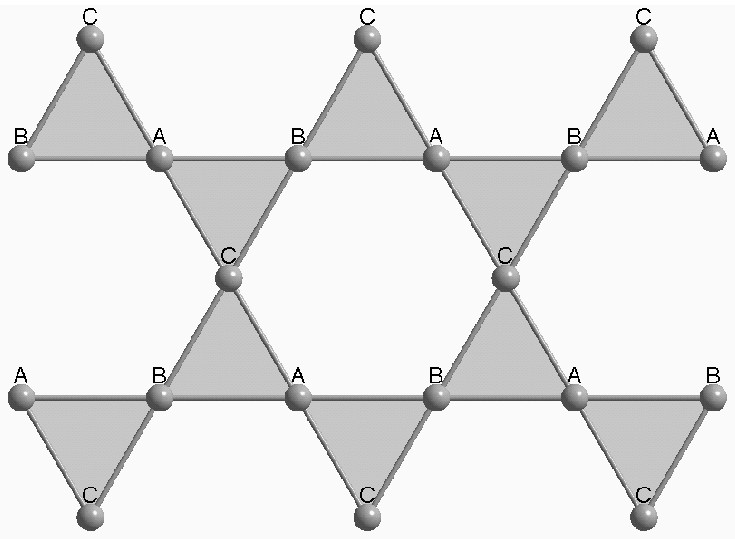}} \subfigure[Pyrochlore lattice]{\includegraphics{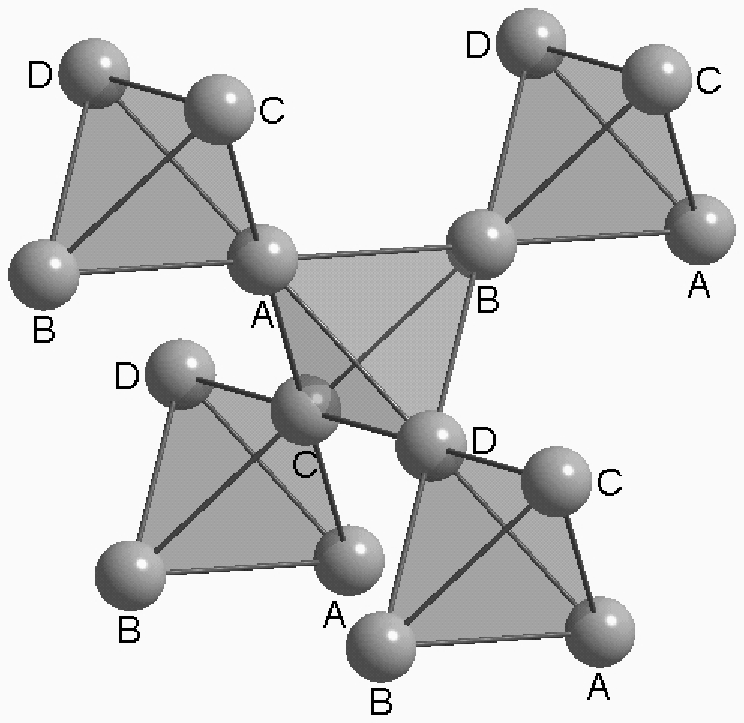}} \subfigure[Linear chain]{\includegraphics{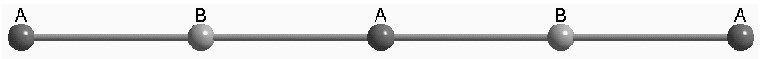}} \par}

\caption{\label{fig.lattices}The lattices we will consider in this work.
\protect\( A,B,C,\ldots \protect \) denote the sublattices in which
the whole lattice is subdivided. All these lattices share the following
property: if we consider a spin inside a cluster formed by \protect\( p\protect \)
spins, there are \protect\( p-1\protect \) NN spins outside the cluster.
Notice also that one spin belonging to one sublattice does not have
NN belonging to the same sublattice.}
\end{figure}

The study of the magnetic ordering in geometrically frustrated antiferromagnets
(GFAF) is not a new problem in Condensed Matter Physics. It was initiated
in the fifties by Wannier, Houtappel, and Anderson\cite{WANNIER1950, HOUTAPPEL1950, ANDERSON1956}.
A further milestone was set by Villain\cite{VILLAIN1979}, who first
argued that the pyrochlore Heisenberg antiferromagnet remains paramagnetic
down to 0 K. However, there has been a renewed interest during recent
years on these lattices, mainly because, experimentally, a rich phenomenology
has been found at low temperatures and novel phases have been identified,
even though these systems are relatively simple crystals\cite{SCHIFFER1996b, Ramirez1994, GINGRAS2000}.
In these materials, the elementary unit of the magnetic structure
is the triangle, which makes it impossible to satisfy all the antiferromagnetic
bonds at the same time, with the result of a macroscopically degenerate
ground state. Among the lattices with the highest degree of frustration
are the pyrochlore and kagome ones. In the former, the magnetic ions
occupy the corners of a 3D arrangement of corner sharing tetrahedra;
in the latter, the magnetic ions occupy the corners of a 2D arrangement
of corner sharing triangles (see Fig.~\ref{fig.lattices}). To summarize
a very long story, it is found that the susceptibility in materials
belonging to this class exhibits a high temperature phase in which
it follows the Curie--Weiss (CW) law. Below the Curie--Weiss temperature,
geometrical frustration inhibits the formation of a long range ordered
(LRO) state, and the system remains paramagnetic, even though there
are strong correlations between units. This phase is universally present
in these systems, and it is called the cooperative paramagnetic phase.
Finally, at a certain temperature, \( T_{f} \), which depends on
the particular material and, usually, is well below the CW temperature,
there appear non universal phases: some of the systems remain paramagnetic
down to the lowest temperature reached experimentally\cite{GARDNER1999},
some of them exhibit non collinear ordered states\cite{RAJU1999, CHAMPION5043},
or even some of them form a spin glass state\cite{MARTINHO2001, WILLS2000, Earle99},
even though the amount of disorder in the structure is very small.
It is thus easy to understand the amount of theoretical attention
these systems have generated\cite{Reimers1991, Moessner99, Moessner98a, Moessner98, MOESSNER2000, Isoda98, Harris92, Canals98, Canals00a, Garanin99, CHUBUKOV1992}.

In the simplest theoretical description of GFAF, the spins on the
lattice are regarded as Heisenberg spins with only nearest neighbor
(NN) interactions. In this picture, it is predicted that the non trivial
degeneracy of the ground state inhibits the formation of a LRO state,
and the system remains paramagnetic down to zero temperature\cite{Reimers1991, MOESSNER2000}.
However, due to the presence of frustration, the NN exchange does
not fix an energy scale on the problem, and any small perturbation
can break the non trivial degeneracy of the ground state and lead
to some kind of ordered state. Therefore, it is especially important
to incorporate these possible perturbations in any model that tries
to explain the low temperature phases of these systems. Examples of
perturbations present in real systems are next nearest neighbor (NNN)
interactions, small anisotropies, long range dipole--dipole interaction,
or dilution by non magnetic impurities, to cite some.

In a series of recent papers\cite{GARCIA-ADEVA2001a, GARCIA-ADEVA2001b, GARCIA-ADEVA2001d},
the present authors have generalized the well known Constant Coupling
method of magnetism\cite{KASTELEJEIN1956, ELLIOTT1960}, to deal with
geometrically frustrated lattices, in both the classical and quantum
cases, in the cooperative paramagnetic region. Also, the connection
between this technique and the so called Phenomenological Renormalization
Group (RG) methods for Ising spins\cite{INDEKEU1982, FITTIPALDI1994, PLASCAK1999}
has been furnished in Ref.~\onlinecite{GARCIA-ADEVA2001c}, where it
was shown that the Generalized Constant Coupling (GCC) method can
be regarded as the zero-th order approximation of one of these RG
techniques, for vectorial or scalar spin systems, the so called, Mean
Field Renormalization Group method, when correlations outside a cluster
are neglected, and the anomalous critical exponent of the system is
taken to be zero. Therefore, it is not so surprising that magnetic
properties calculated within the framework of the classical GCC method
were found to be in excellent agreement with Monte Carlo calculations
for those same quantities\cite{Chalker92a, Reimers1993b, Moessner99},
for both the kagome and pyrochlore lattices. The main idea behind
this technique is very simple, and it is based in the experimental
and numerical observation that correlations in GFAF are always short
ranged\cite{Canals98, Canals00a, Moessner98a, Moessner99, GARDNER1999, RAJU1999}.
Therefore, in order to calculate any thermodynamic quantity, we can
start by calculating the partition function of an isolated cluster
with \( p \) spins and, later, add the interactions with the surrounding
\( p-1 \) spins outside the cluster (see Fig.~\ref{fig.lattices})
by means of an effective field which is fixed by a self-consistency
condition. This, almost phenomenological, formulation of the method,
allows one to easily compute thermodynamic quantities, mainly the
susceptibility, in the paramagnetic region, by using a single effective
field to characterize all the interactions with spins outside the
cluster. However, it is not very useful to study the critical properties
of these systems, except for ferromagnetic interactions, which is
not as interesting as the antiferromagnetic case, due to the absence
of geometrical frustration (for Heisenberg spins). The reason is very
simple. It is now generally accepted that the ordered state of these
GFAF, if any, is not a simple antiferromagnetic configuration, as
in bipartite lattices, but a non-collinear (NC) one, in which the
total spin momentum of the unit is identically zero\cite{VILLAIN1979, Reimers1991}.
Obviously, we cannot characterize such an ordered state by using a
single order parameter fixed by a single effective field (EF). What
is even worse, even we know that we have to introduce different order
parameters for each of the spins in the cluster, and different (EF)
for the neighboring spins outside the cluster, it is not easy to intuitively
see how these EF enter into the expressions of the order parameters.
Therefore, we need to go beyond the phenomenological formulation of
the GCC method introduced in the aforementioned works. This is precisely
the intention of this work. We construct the GCC method from first
principles, by using rigorous identities in terms of averages over
finite size clusters. In this way, the EF naturally emerge as averages
over the microscopic spin variables. This microscopic formulation
of the GCC method not only allows us to recover all the results obtained
Refs.~\onlinecite{GARCIA-ADEVA2001a} and \onlinecite{GARCIA-ADEVA2001d}
in the paramagnetic regime but, also, gives us the possible ordered
configurations of the system, and the conditions for a transition
to one of these states to occur. Moreover, as the GCC method is constructed
in the real space, it is especially easy to include further perturbations
in the system. Particularly, we study the effect of next nearest neighbor
(NNN) interactions and site dilution by non-magnetic impurities.

Apart from the microscopic formulation of the GCC method itself, we
present various interesting results in this work. We find the possible
ordered configurations for systems formed by corner sharing units.
These systems have the interesting property that the corresponding
lattices are formed by units with \( p=d+1 \) spins (\( d \) the
dimensionality), and each of the spins in the unit has \( p-1 \)
NN outside the unit. Thus, for \( d=3 \), we have the pyrochlore
lattice; for \( d=2 \), the kagome lattice, and for \( d=1 \), the
linear chain. We have studied this last case for the sake of completeness,
even though this system is not frustrated. It is found that, apart
from the trivial ferromagnetic configuration for ferromagnetic interactions,
the ordered state would correspond to a non-collinear configuration
given by the condition that the total magnetization of the unit is
zero (for the especial case of the linear chain this corresponds to
standard antiferromagnetic order). For NN interactions only, there
is no transition to such an ordered state at any finite temperature,
for none of the systems considered, as it is now generally accepted\cite{Reimers1991, MOESSNER2000}.
However, when NNN interactions are taken into account, there is a
transition to such a NC ordered state for some ranges of the coupling
parameters. We have only studied the pyrochlore lattice in this case,
as we do not think the results of the present method for the critical
behavior of these systems are reliable in dimension lower than 3 (actually,
the results of any model based on the idea of finite size clusters),
as the long wavelength fluctuations around the NC ordered state are
not properly taken into account, and they are very likely to be specially
important in 2 dimensions (kagome lattice). Also, the effect of site
dilution is considered in the present work. It is found that, in this
framework, dilution by itself is not enough to stabilize a NC ordered
state. Finally, we also consider both dilution and NNN together, and
the phase diagrams for different amounts of dilution are calculated.

The remaining of the paper is organized as follows: In the following
Section, we introduce the systems studied in this work, the Hamiltonian
of these systems, the notation, and the self-consistency condition
that fixes the EF. In Section \ref{sec.order.parameters} we present
the calculation of the order parameters for finite cluster of size
1 and \( p \) (\( p=4,3 \), and \( 2 \) for the pyrochlore, kagome,
and linear chain, respectively). The connection with the original
formulation of the GCC method is furnished in Section \ref{sec.connection}.
In Section \ref{sec.critical} we study the conditions for a transition
to a long range ordered state. In Section \ref{sec.nnn} we include
NNN interactions and study how this additional interaction modifies
the conclusions of Section \ref{sec.critical}. In Section \ref{sec.dilution}
we develop the formalism to deal with site dilution by non-magnetic
impurities, and apply it to the present problem. In Section \ref{sec.nnn.dilution}
we consider simultaneously the effect of both NNN interactions and
site dilution. Section \ref{sec.conclusions} is devoted to the discussion
of the results and the conclusions. Finally, we have included two
appendixes to clarify some of the most involved details of the evaluation
of the partition function of the different clusters subject to inhomogeneous
EF.

\section{Preliminaries}

As stated above, the original formulation of the GCC method\cite{GARCIA-ADEVA2001a, GARCIA-ADEVA2001b}
is not well suited to study the critical behavior of GFAF. Indeed,
it is now generally accepted that the ground state of the pyrochlore
and kagome lattices is fixed by the condition that the total spin
momentum of the tetrahedral or triangular unit, respectively, is zero,
though frustration inhibits the formation of such a state, and the
system remains paramagnetic in the whole temperature range, if only
NN interactions are present. Obviously, we cannot characterize such
an ordered state with a single order parameter, as we did in Refs.~\onlinecite{GARCIA-ADEVA2001a} and
\onlinecite{GARCIA-ADEVA2001b}, which restricts the applicability
of the GCC method to the study of the paramagnetic region or, at most,
the ferromagnetic case, which is of less interest, due to the absence
of geometrical frustration. A relatively simple way of circumventing
this shortcoming of the method has been pointed out in Refs.~\onlinecite{Reimers1991}
and \onlinecite{GARCIA-ADEVA2001c}, and essentially consists on introducing
different sublattices characterized by different order parameters.
The minimum number of sublattices we need to introduce is given by
the number of spins in the cluster, that is, 3 and 4 for the kagome
and pyrochlore lattices, respectively (see Fig.~\ref{fig.lattices}).
With this subdivision, spins belonging to one sublattice only interact
with spins belonging to different sublattices. Therefore, the dimensionless
Heisenberg Hamiltonian of a lattice formed by corner sharing clusters
with \( p \) spins, and only NN interactions, in the presence of
a uniform magnetic field can be put in the form\begin{equation}
\mathcal{H}=K\sum _{\alpha \neq \beta }\sum _{\left\langle i,j\right\rangle }\vec{s}_{i\alpha }\cdot \vec{s}_{j\beta }+\vec{H}_{0}\cdot \sum _{\alpha ,i}\vec{s}_{i\alpha },
\end{equation}
 where \( \vec{s}_{i\alpha } \) are the classical Heisenberg spins
of unit length. The \( \alpha  \) index labels the sublattice to
which the considered spin belongs (and takes the values \( \alpha =A,B,C,\ldots  \)),
whereas the \( i \) index labels the spins belonging to a given sublattice.
\( K=\frac{J}{T} \) (\( K>0 \) for ferromagnetic interactions and
\( K<0 \) for antiferromagnetic ones) and \( \vec{H}_{0}=\frac{\vec{h}_{0}}{T}, \)
with \( \vec{h}_{0} \) the applied magnetic field.

The central idea of the GCC method is to calculate the order parameter
(for each sublattice) for two clusters of different sizes, \( p \)
and \( p' \), replacing the effect of the spins not included in the
cluster by fixed effective fields, which act as a symmetry breaking
field (SBF) and, by using the self-consistency condition\begin{equation}
\label{scaling.relation}
\vec{m}_{p'}(K)=\vec{m}_{p}(K),
\end{equation}
 obtain an equation to determine these effective fields and, thus,
the magnetization. The order parameter is calculated by making use
of the Callen-Suzuki identity\cite{SUZUKI1965, GARCIA-ADEVA2001c}\begin{equation}
\vec{m}_{p\alpha }=\left\langle \vec{s}_{i\alpha }\right\rangle =\left\langle \frac{Tr_{p}\vec{s}_{i\alpha }e^{\mathcal{H}_{p}}}{Tr_{p}e^{\mathcal{H}_{p}}}\right\rangle _{\mathcal{H}},
\end{equation}
 where the partial trace is taken over the set of \( p \) variables
specified by the finite size cluster Hamiltonian \( \mathcal{H}_{p} \);
\( \vec{m}_{\alpha } \) is the corresponding order parameter, and
\( \left\langle \ldots \right\rangle _{\mathcal{H}} \) indicates
the usual canonical thermal average over the ensemble defined by the
complete Hamiltonian \( \mathcal{H} \).

\section{Evaluation of the order parameters in the presence of the SBF}

\label{sec.order.parameters}

\subsection{1-spin cluster order parameter}

\begin{figure}[htb!]
{\centering \subfigure[Kagome lattice]{\includegraphics{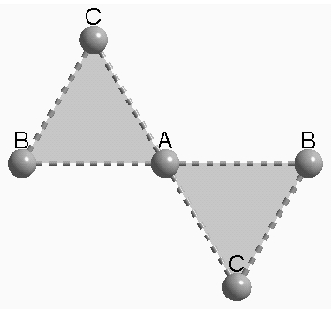}} \subfigure[Pyrochlore lattice]{\includegraphics{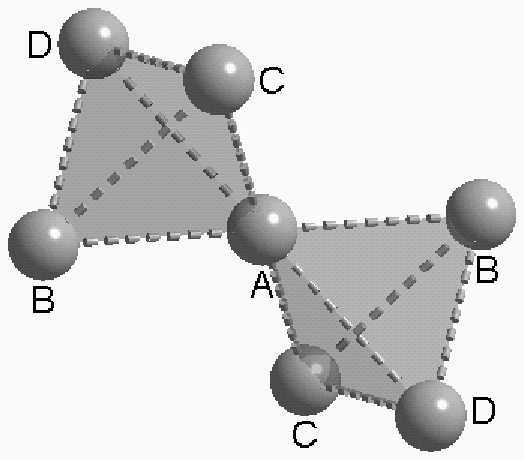}} \par}

{\centering \subfigure[Linear chain]{\includegraphics{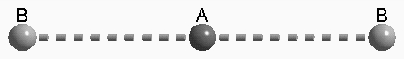}} \par}

\caption{\label{fig.1spin.clusters}1-spin clusters for the different sublattices
considered in this work. The interactions with spins outside the cluster,
i.e., those creating the EF, have been represented by dashed bonds.
Notice that there are 2 NN on each sublattice, different from the
one to which the considered spin belongs, outside the cluster.}
\end{figure}

Let us consider the cluster defined by the spin 1 in sublattice \( \alpha  \)
(see Fig.~\ref{fig.1spin.clusters}). The Hamiltonian of this cluster
is given by\begin{equation}
\mathcal{H}_{1\alpha }=\vec{s}_{1\alpha }\left( K\sum _{\beta \neq \alpha }\sum _{i}^{z}\vec{s}_{i\beta }+\vec{H}_{0}\right) =\vec{s}_{1\alpha }\cdot \vec{\xi }_{1\alpha },
\end{equation}
 where the subindex \( 1\alpha  \) makes reference to the fact that
the Hamiltonian corresponds to a cluster of 1 spin belonging to sublattice
\( \alpha  \), and \( \vec{\xi }_{1\alpha } \) stands for the symmetry
breaking field acting on the spin of the 1-spin cluster, which belongs
to sublattice \( \alpha  \). The sum over the \( i \) index is performed
over the NN of the \( 1\alpha  \) spin, with \( z \) the number
of NN in each of the sublattices (\( z=2 \) for both the kagome and
pyrochlore lattices).

By using the Callen--Suzuki identity we can calculate the order parameter
in sublattice \( \alpha  \) for a cluster formed by 1 spin\begin{equation}
\vec{m}_{1\alpha }=\left\langle \vec{s}_{1\alpha }\right\rangle =\left\langle \frac{Tr_{1\alpha }\vec{s}_{1\alpha }e^{\mathcal{H}_{1\alpha }}}{Tr_{1\alpha }e^{\mathcal{H}_{1\alpha }}}\right\rangle _{\mathcal{H}}.
\end{equation}
 As commented above, \( Tr_{1\alpha } \) represents the partial trace
with respect to the variables of the spin \( 1\alpha  \), and the
symbol \( \left\langle \ldots \right\rangle _{\mathcal{H}} \) stands
for the usual canonical thermal average taken over the ensemble defined
by the complete Hamiltonian \( \mathcal{H} \). This equation can
be put in the more convenient form\begin{equation}
\vec{m}_{1\alpha }=\left\langle \nabla _{\vec{\xi }_{1\alpha }}\ln Tr_{1\alpha }e^{\mathcal{H}_{1\alpha }}\right\rangle _{\mathcal{H}}.
\end{equation}
For classical vectorial spins of fixed length, the trace operation
is given by\begin{equation}
Z_{1\alpha }(K)=Tr_{1\alpha }e^{\vec{s}_{1\alpha }\cdot \vec{\xi }_{1\alpha }}=\int d\Omega \, e^{\vec{s}\cdot \vec{\xi }_{1\alpha }},
\end{equation}
 where the integration is performed over the angular degrees of freedom
of the spin. This integral is trivially evaluated to give\begin{equation}
Z_{1\alpha }(K)=2\frac{\sinh \xi _{1\alpha }}{\xi _{1\alpha }}
\end{equation}
 and, taking into account the well known relation \( \nabla _{\vec{r}}r=\widehat{n} \),
where \( \widehat{n} \) is a unitary vector in the direction of \( \vec{r} \),
we can easily calculate the order parameter for the 1-spin cluster
as\begin{equation}
\vec{m}_{1\alpha }=\left\langle L(\xi _{1\alpha })\right\rangle \, \widehat{n}_{1\alpha },
\end{equation}
 where \( L(x) \) is the Langevin function\cite{SMART1966} and \( \widehat{n}_{1\alpha } \)
is an unitary vector in the direction of the corresponding SBF. 

Near a critical point, as we expect the SBF to be very small, we can
expand the previous expression and keep only linear terms in the SBF\begin{equation}
\vec{m}_{1\alpha }\simeq \frac{\left\langle \vec{\xi }_{1\alpha }\right\rangle }{3}.
\end{equation}
 By defining the effective fields \( \vec{h}'_{\alpha }=J\left\langle \vec{s}_{i\alpha }\right\rangle _{\mathcal{H}}=J\left\langle \vec{s}_{j\alpha }\right\rangle _{\mathcal{H}} \)
(\( i,j\neq 1 \)), we arrive to the final expression for the order
parameter\begin{equation}
\label{order.param.1spin}
\vec{m}_{1\alpha }=\frac{2}{3T}\sum _{\beta \neq \alpha }\vec{h}'_{\beta }+\frac{\vec{h}_{0}}{3T},
\end{equation}
 where we have taken into account that each spin has 2 NN on each
sublattice different from the considered one. For example, for the
pyrochlore lattice with no applied magnetic field we have \begin{equation}
\vec{m}_{1\alpha }=\frac{2}{3T}(\vec{h}'_{\beta }+\vec{h}'_{\gamma }+\vec{h}'_{\delta }),
\end{equation}
 where \( \alpha  \), \( \beta  \), \( \gamma  \), and \( \delta  \)
can take the values \( A \), \( B \), \( C \), or \( D \). For
the kagome lattice\[
\vec{m}_{1\alpha }=\frac{2}{3T}(\vec{h}'_{\beta }+\vec{h}'_{\gamma }),\]
 where \( \alpha  \), \( \beta  \), and \( \gamma  \) can take
the values \( A \), \( B \), or \( C \). For a cluster with 2 spins,
we have a linear chain, for which \begin{equation}
\vec{m}_{1\alpha }=\frac{2}{3T}\vec{h}'_{\beta },
\end{equation}
 where \( \alpha  \) and \( \beta  \) can take the values \( A \)
or \( B \).

Another interesting point is that if we keep all the terms in the
expansion of the order parameter, and make use of the decoupling approximation
\begin{equation}
\label{decoupling}
\left\langle s_{i\alpha }s_{j\beta }\ldots s_{k\gamma }\right\rangle _{\mathcal{H}}\simeq \left\langle s_{i\alpha }\right\rangle _{\mathcal{H}}\left\langle s_{j\beta }\right\rangle _{\mathcal{H}}\ldots \left\langle s_{k\gamma }\right\rangle _{\mathcal{H}},
\end{equation}
 and further assume \( \vec{h}'_{\alpha }=\vec{m}_{1\alpha } \),
we recover the usual self consistent Curie--Weiss equation for the
order parameter\begin{equation}
\vec{m}_{1\alpha }=L\left( 2\, K\sum _{\beta \neq \alpha }\vec{m}_{1\beta }+\vec{H}_{0}\right) .
\end{equation}
 However, this expression would predict a transition to an ordered
state (for antiferromagnetic interactions) at a finite temperature
given by \( T_{N}=-\Theta _{CW}=\frac{2(p-1)\left| J\right| }{3} \),
which it is well known to be incorrect for GFAF. That is one of the
reasons why constructing the effective fields by the self-consistency
condition \eqref{scaling.relation} provides a much better theory for
GFAF, as we will see below.

\subsection{\protect\( p\protect \)-spin cluster order parameter}

\begin{figure}[htb!]
{\centering \subfigure[Kagome lattice]{\includegraphics{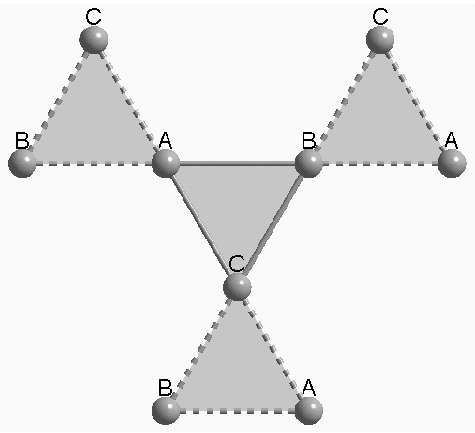}} \subfigure[Pyrochlore lattice]{\includegraphics{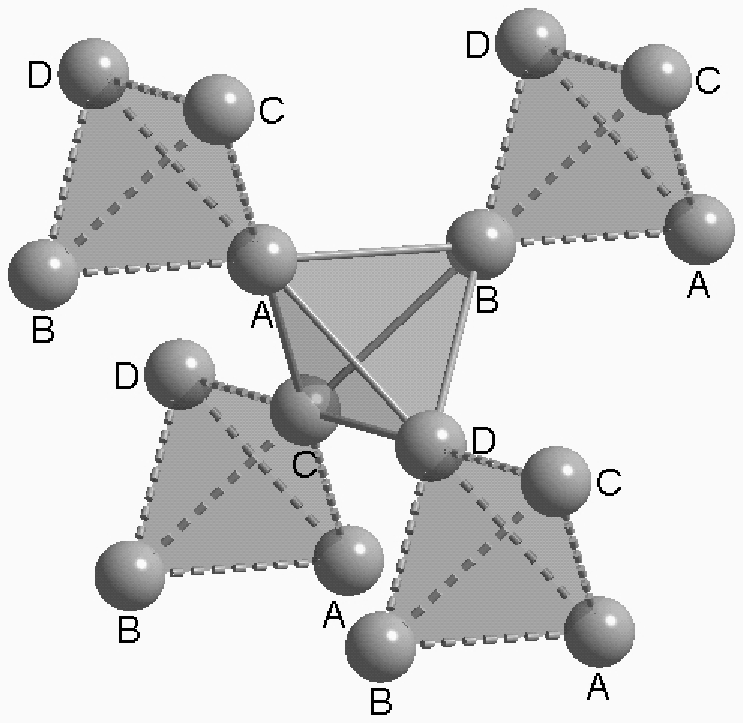}} \par}

{\centering \subfigure[Linear chain]{\includegraphics{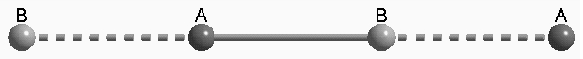}} \par}

\caption{\label{fig.pspin.clusters}\protect\( p\protect \)-spin clusters
for the lattices considered in this work. Again, we have represented
the interactions with the spins creating the EF by dashed bonds. Notice
that there is 1 spin outside the cluster on each sublattice different
from the considered spin of the cluster.}
\end{figure}

The calculation of the order parameter for a cluster with \( p \)
spins in the presence of the SBF created by the neighboring spins
is a highly non-trivial problem, as we will see in the following.
The size of the clusters is chosen so that we consider the smallest
number of spins that include all the different sublattices. That is,
\( p=3 \) for a kagome lattice (a triangular cluster), \( p=4 \)
for a pyrochlore (a tetrahedral cluster), etc (see Fig.~\ref{fig.pspin.clusters}).
The Hamiltonian for such a cluster can be put as\begin{equation}
\mathcal{H}_{p}=K\sum _{\alpha \neq \beta }\vec{s}_{1\alpha }\cdot \vec{s}_{1\beta }+\sum _{\alpha }\vec{s}_{1\alpha }\cdot \vec{\xi }_{p\alpha }.
\end{equation}
In this Hamiltonian, the first term represents the interaction between
the spins inside the cluster of size \( p \), whereas the second
term represents the interaction of those same spins with the symmetry
breaking fields \begin{equation}
\vec{\xi }_{p\alpha }=K\sum _{\beta \neq \alpha }\sum _{i}^{z-1}\vec{s}_{i\beta }+\vec{H}_{0},
\end{equation}
 created by the \( p-1 \) spins outside the cluster. The sum over
\( i \) goes over all the NN spins except the ones already included
in the cluster.

The order parameter for each of the sublattices included in the cluster,
obtained from the \( p \)-spin cluster can be now put as\begin{equation}
\vec{m}_{p\alpha }=\left\langle \frac{Tr_{1A,1B,1C\ldots }\vec{s}_{1\alpha }e^{\mathcal{H}_{p}}}{Tr_{1A,1B,1C\ldots }e^{\mathcal{H}_{p}}}\right\rangle _{\mathcal{H}},
\end{equation}
 or, more conveniently\begin{equation}
\label{order2}
\vec{m}_{p\alpha }=\left\langle \nabla _{\vec{\xi }_{p\alpha }}\ln Tr_{1A,1B,1C\ldots }e^{\mathcal{H}_{p}}\right\rangle _{\mathcal{H}}.
\end{equation}
 In these expressions, \( Tr_{1A,1B,1C\ldots } \) represents the
trace over the spin degrees of freedom in the \( p \)-spin cluster.
The reduced trace operation for Heisenberg spins is given by\begin{equation}
\label{partition.cluster}
Z_{p}(K)=\int \prod _{\alpha }d\Omega _{\alpha }e^{\mathcal{H}_{p}}
\end{equation}
 It is impossible to find a closed expression for this partition function
of the cluster in the presence of inhomogeneous SBF (the case with
no SBF has been studied by Moessner and Berlinsky in Ref.~\onlinecite{Moessner99}).
However, as we will be mainly interested in the region near the critical
point (if any) in the absence of applied field, where the SBF are
expected to be small, and in the paramagnetic region, where the SBF
will be proportional to the applied field and, again, small, it is
enough for our purposes to calculate reduced partition function up
to second order in the SBF. The main steps of the calculation are
presented in Appendix \ref{appendix1}, and the result is given by
\begin{equation}
\label{partition.function.cluster}
Z_{p}(K)\simeq z_{p}^{0}(K)+\frac{z_{p}^{0}(K)}{6}\sum _{\alpha }\xi _{p\alpha }^{2}-\frac{z_{p}^{1}(K)}{6}\sum _{\alpha \neq \beta }\vec{\xi }_{p\alpha }\cdot \vec{\xi }_{p\beta },
\end{equation}
 for \( p>1 \), where \begin{equation}
z^{0}_{p}(K)=\int _{0}^{\infty }dq\, q^{2}\, e^{\frac{q^{2}}{2K}}\left( \frac{\sin q}{q}\right) ^{p}
\end{equation}
 and\begin{equation}
z_{p}^{1}(K)=\int _{0}^{\infty }dq\, q^{2}\, e^{\frac{q^{2}}{2K}}\left( \frac{\sin q}{q}\right) ^{p-2}\left( \frac{\cos q}{q}-\frac{\sin q}{q^{2}}\right) ^{2}.
\end{equation}
 Expressions for these functions are given in Appendix \ref{appendix2}.
Incidentally, the first term in \eqref{partition.function.cluster} corresponds
to the partition function for non-interacting clusters with no applied
field, which were first calculated by Moessner and Berlinsky\cite{Moessner99}.
The additional terms in \eqref{partition.function.cluster} allow us
to study not only the paramagnetic regime, but also the behavior of
the system near a critical point, as we will see below. 

Once we have this reduced partition function we can calculate the
order parameter up to first order in the SBF by taking into account
\eqref{order2}\begin{equation}
\vec{m}_{p\alpha }=\frac{\left\langle \vec{\xi }_{p\alpha }\right\rangle }{3}-\frac{z_{p}^{1}(K)}{3z_{p}^{0}(K)}\sum _{\beta \neq \alpha }\left\langle \vec{\xi }_{p\beta }\right\rangle .
\end{equation}
 Taking into account the definition of the SBF, and the fact that
\( J\left\langle \vec{s}_{i\alpha }\right\rangle _{\mathcal{H}}=J\left\langle \vec{s}_{j\alpha }\right\rangle _{\mathcal{H}}=\vec{h}'_{\alpha } \),
after a little algebra we arrive at\begin{equation}
\label{order.param.pspin}
\vec{m}_{p\alpha }=\frac{1}{3T}\left\{ \left[ 1-(p-1)a_{p}(K)\right] \vec{h}_{0}+\left( 1-(p-2)a_{p}(K)\right) \sum _{\beta \neq \alpha }\vec{h}'_{\beta }-(p-1)a_{p}(K)\vec{h}'_{\alpha }\right\} ,
\end{equation}
 where we have introduced the function\begin{equation}
a_{p}(K)=\frac{z_{p}^{1}\left( K\right) }{z_{p}^{0}\left( K\right) }.
\end{equation}

\section{Calculation of the susceptibility: Connection with the original formulation
of the GCC method}

\label{sec.connection}

In the following, we will find more useful to express all the thermodynamic
quantities in terms of the dimensionless temperature \begin{equation}
\widetilde{T}=\frac{1}{\left| K\right| }=\frac{T}{\left| J\right| }.
\end{equation}
 In terms of this parameter, we have\begin{equation}
\label{new.order.param.1spin}
\vec{m}_{1\alpha }=\frac{1}{3\left| J\right| \widetilde{T}}\left( \vec{h}_{0}+2\sum _{\beta \neq \alpha }\vec{h}'_{\beta }\right) ,
\end{equation}
 and\begin{equation}
\vec{m}_{p\alpha }=\frac{1}{3\left| J\right| \widetilde{T}}\left\{ \left[ 1-(p-1)A_{p}(\widetilde{T})\right] \vec{h}_{0}+\left( 1-(p-2)A_{p}(\widetilde{T})\right) \sum _{\beta \neq \alpha }\vec{h}'_{\beta }-(p-1)A_{p}(\widetilde{T})\vec{h}'_{\alpha }\right\} ,
\end{equation}
 where\begin{equation}
A_{p}(\widetilde{T})=a_{p}(\widetilde{T})
\end{equation}
 for ferromagnetic interactions and\begin{equation}
A_{p}(\widetilde{T})=a_{p}(-\widetilde{T})
\end{equation}
 for antiferromagnetic ones. The self consistency condition (see eq.
\eqref{scaling.relation})\begin{equation}
\vec{m}_{1\alpha }=\vec{m}_{p\alpha }
\end{equation}
 leads to the following system of linear equations\begin{equation}
\label{general.system.eq}
(p-1)\, A_{p}(\widetilde{T})\, \vec{h}'_{\alpha }+\left[ 1+(p-2)\, A_{p}(\widetilde{T})\right] \sum _{\beta \neq \alpha }\vec{h}'_{\beta }=-(p-1)\, A_{p}(\widetilde{T})\, \vec{h}_{0},
\end{equation}
 that has the solution\begin{equation}
\vec{h}'_{\alpha }=\vec{h}'=-\frac{A_{p}(\widetilde{T})}{\left[ 1+(p-1)A_{p}(\widetilde{T})\right] }\, \vec{h}_{0}.
\end{equation}
 Substituting back in \eqref{new.order.param.1spin} we find\begin{equation}
\vec{m}_{p\alpha }=\frac{1}{3\left| J\right| \widetilde{T}}\, \frac{1-(p-1)A_{p}(\widetilde{T})}{1+(p-1)A_{p}(\widetilde{T})}\, \vec{h}_{0},
\end{equation}
 that is, the susceptibility is given by\begin{equation}
\label{suscep.nnn}
\chi _{p}^{gcc}=\frac{1}{3\left| J\right| \widetilde{T}}\, \frac{1-(p-1)A_{p}(\widetilde{T})}{1+(p-1)A_{p}(\widetilde{T})}.
\end{equation}
 By direct substitution, it is simple to verify the relation\begin{equation}
\varepsilon _{p}(\widetilde{T})=-(p-1)\, A_{p}(\widetilde{T}),
\end{equation}
 where \( \varepsilon _{p}(\widetilde{T}) \) is defined as\cite{GARCIA-ADEVA2001a}
\begin{equation}
\varepsilon _{p}(\widetilde{T})=\frac{2\widetilde{T}^{2}}{p}\frac{\partial }{\partial \widetilde{T}}\ln z_{p}^{0}(\widetilde{T})-1
\end{equation}
 so we can finally put\begin{equation}
\chi _{p}^{gcc}=\frac{1}{3\left| J\right| \widetilde{T}}\, \frac{1+\varepsilon _{p}(\widetilde{T})}{1-\varepsilon _{p}(\widetilde{T})},
\end{equation}
 in complete agreement with the result in Ref.~\onlinecite{GARCIA-ADEVA2001a}.

\section{Calculation of the critical points}

\label{sec.critical}

In the absence of applied field, the system of equations \eqref{general.system.eq}
takes the simpler form\begin{equation}
\label{homo.general.system.eq}
(p-1)\, A_{p}(\widetilde{T})\, \vec{h}'_{\alpha }+\left[ 1+(p-2)A_{p}(\widetilde{T})\right] \sum _{\beta \neq \alpha }\vec{h}'_{\beta }=\vec{0},
\end{equation}
and it has two different non-trivial solutions. The first one occurs
if \begin{equation}
A_{p}(\widetilde{T})=1
\end{equation}
 or, equivalently,\begin{equation}
\varepsilon _{p}(\widetilde{T})=-(p-1),
\end{equation}
and corresponds to an ordered state in which\begin{equation}
\label{condition.nc}
\sum _{\alpha }\vec{m}_{\alpha }=\vec{0}.
\end{equation}
 We will call this kind of state a non-collinear (NC) ordered state.
Actually, this condition includes, as particular cases, states in
which spins in different sublattices are antiferromagnetically aligned
with each other, that is, collinear states. However, the most general
state described by this condition is a non-collinear one. The second
solution of \eqref{homo.general.system.eq} occurs if \begin{equation}
A_{p}(K)=-\frac{1}{p-1}
\end{equation}
 or, equivalently, \begin{equation}
\varepsilon _{p}(\widetilde{T})=1,
\end{equation}
and corresponds to an ordered state in which \begin{equation}
\vec{m}_{\alpha }=\vec{m}_{\beta }=\vec{m}_{\gamma }=\ldots \, \, \, \, \, \, \, \, \, \forall \alpha ,\beta ,\gamma ,\ldots ,
\end{equation}
 that is, ferromagnetic order.
\begin{figure}[htb!]
{\centering \subfigure[\label{fig.noncollinear}Non-collinear order]{\includegraphics{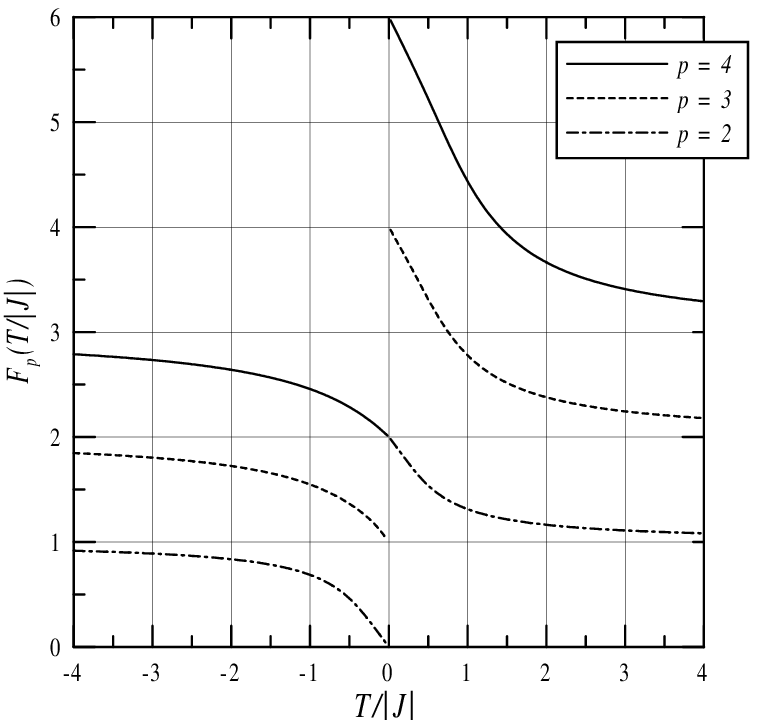}} \subfigure[\label{fig.ferro}Ferromagnetic order]{\includegraphics{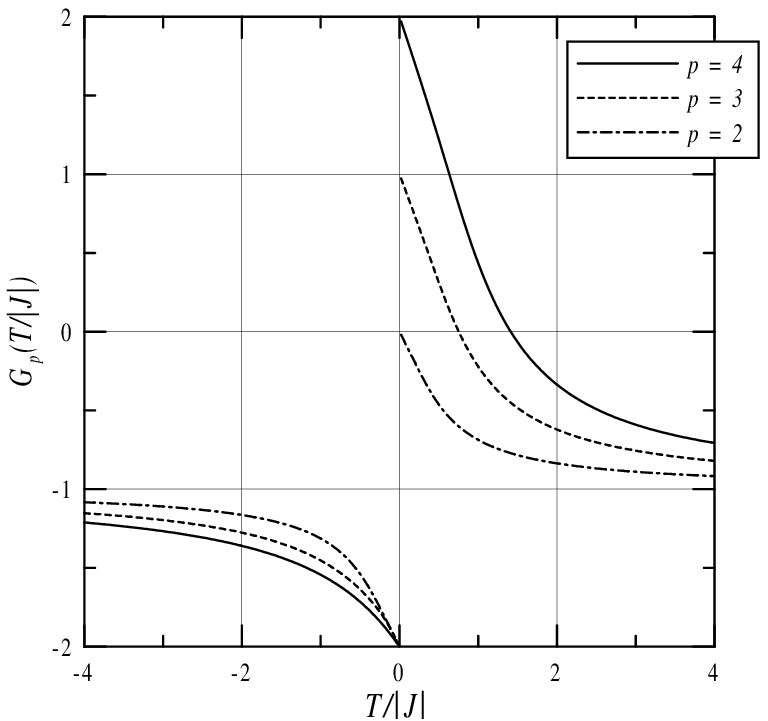}} \par}

\caption{Graphical representation of the conditions for (a) Non-collinear
order and (b) Ferromagnetic order.}
\end{figure}

The previous condition for the existence at any finite temperature
of a ferromagnetic ordered state is equivalent to studying if the
function \( G_{p}(\widetilde{T})=\varepsilon _{p}(\widetilde{T})-1 \)
changes its sign for any finite value of \( \widetilde{T} \). This
function has been depicted in Fig.~\ref{fig.ferro} for the pyrochlore
lattice, kagome lattice, and linear chain. We can see that the present
method correctly predicts the existence of a transition for the pyrochlore
lattice for a value finite value of \( \widetilde{T} \) (\( K_{c}=0.71 \)),
and no finite \( T \) transition for the linear chain, but incorrectly
predicts a transition for the kagome lattice (\( K_{c}=1.33 \)).
This is due to the fact that the long wavelength fluctuations around
the ferromagnetic state are not taken into account in the present
method, and these fluctuations destroy the ordered state in 2 dimensions.
On the other hand, in the linear chain case, even short range correlations,
properly described in the model, are enough to destroy any form of
long range order.

The condition for the existence of a non-collinear ordered state is
equivalent to study if the function \( F_{p}(\widetilde{T})=\varepsilon _{p}(\widetilde{T})+(p-1) \)
changes its sign for any finite value of \( \widetilde{T} \). This
function has been depicted in Fig.~\ref{fig.noncollinear}. We can
see that the model correctly predicts that there is no transition
to an ordered state for any of the systems considered.

For comparison, we have also applied the present GCC method to the
Heisenberg model with NN interactions on a simple cubic (s.c.) lattice.
In the framework of the CW theory, it is predicted that both the s.c.
and the pyrochlore lattices should have the same transition temperature
to a LRO state, as they have the same number of NN. For the s.c.,
it is enough, for our purposes, to consider 2 different sublattices.
Each spin in one sublattice has 6 NN in the other sublattice. Once
we take into account this difference with respect to the pyrochlore
lattice (which has \( p-1 \) NN, as mentioned above), we obtain that,
for ferromagnetic NN interactions, there is a transition to a ferromagnetic
state for \( K_{c}=0.615 \), to be compared with more refined estimates
of this quantity from high-precision MC calculations\cite{CHEN1993},
\( K_{c}=0.693 \). For antiferromagnetic interactions, a transition
to an antiferromagnetic (AF) state occurs at \( K_{c}=-0.615 \).
Therefore, we can see that the present method is able to distinguish
between the different topologies of the lattices, even when they have
the same number of NN. Furthermore, the estimation of the critical
point for the s.c. is very good (up to \textasciitilde{}10\%), even
though we have considered the smallest possible cluster that can describe
the AF ordered state of the s.c. lattice.

As mentioned in the Introduction, one consequence of the presence
of geometrical frustration in the system is that the NN exchange does
not define an energy scale in the problem. Therefore, any perturbation
is a strong perturbation, and can break the non-trivial degeneracy
of the ground state, selecting one ordered state. In the following
sections, we will study if NNN interactions and site dilution by non-magnetic
impurities, which are always present in real systems, really break
this non-trivial degeneracy, that is, if these effects are able to
select an ordered state at finite temperature. Furthermore, we will
focus the discussion on the pyrochlore lattice, for the reasons explained
in the Introduction.

\section{Effect of NNN interactions}

\label{sec.nnn}

In order to include the effect of NNN interactions, we only need to
redefine the SBF in the following way\begin{equation}
\left\langle \vec{\xi }_{1\alpha }\right\rangle =\vec{H}_{0}+K_{1}\sum _{\beta \neq \alpha }\sum _{i}^{z}\left\langle \vec{s}_{i\beta }\right\rangle +K_{2}\sum _{\beta \neq \alpha }\sum _{i}^{z'}\left\langle \vec{s}_{i\beta }\right\rangle =\frac{1}{3\left| J_{1}\right| \widetilde{T}}\left[ \vec{h}_{0}+2(1+z'\, \lambda )\sum _{\beta \neq \alpha }\vec{h}'_{\beta }\right] ,
\end{equation}
 where \( z' \) stands for the number of NNN on each sublattice (\( z'=2 \)
for the kagome lattice and \( z'=4 \) for the pyrochlore lattice)
and \( \lambda =\frac{J_{2}}{J_{1}} \), with \( J_{1} \) the NN
exchange coupling and \( J_{2} \) the NNN exchange coupling. For
the cluster with \( p \) spins\begin{equation}
\left\langle \vec{\xi }_{p\alpha }\right\rangle =\vec{H}_{0}+K_{1}\sum _{\beta \neq \alpha }\sum _{i}^{z-1}\left\langle \vec{s}_{i\beta }\right\rangle +K_{2}\sum _{\beta \neq \alpha }\sum _{i}^{z'}\left\langle \vec{s}_{i\beta }\right\rangle =\frac{1}{3\left| J_{1}\right| \widetilde{T}}\left[ \vec{h}_{0}+(1+z'\, \lambda )\sum _{\beta \neq \alpha }\vec{h}'_{\beta }\right] ,
\end{equation}
 and\begin{equation}
\sum _{\beta \neq \alpha }\left\langle \vec{\xi }_{p\beta }\right\rangle =\frac{1}{3\left| J_{1}\right| \widetilde{T}}\left\{ (p-1)\vec{h}_{0}+(1+z'\, \lambda )\left[ (p-1)\vec{h}'_{\alpha }+(p-2)\sum _{\beta \neq \alpha }\vec{h}'_{\beta }\right] \right\} .
\end{equation}
 Therefore, the self-consistency condition that fixes the effective
fields is given by\begin{equation}
\label{nnn.system.eq}
(1+z'\, \lambda )(p-1)A_{p}(\widetilde{T})\vec{h}'_{\alpha }+\left[ 1+(1+z'\, \lambda )(p-2)A_{p}(\widetilde{T})\right] \sum _{\beta \neq \alpha }\vec{h}'_{\beta }=-(p-1)A_{p}(\widetilde{T})\vec{h}_{0},
\end{equation}
 and its solution is given by\begin{equation}
\vec{h}'_{\alpha }=-\frac{A_{p}(\widetilde{T})}{\left[ 1+(1+z'\, \lambda )(p-1)A_{p}(\widetilde{T})\right] }\, \vec{h}_{0}.
\end{equation}

\begin{figure}[htb!]
{\centering \includegraphics{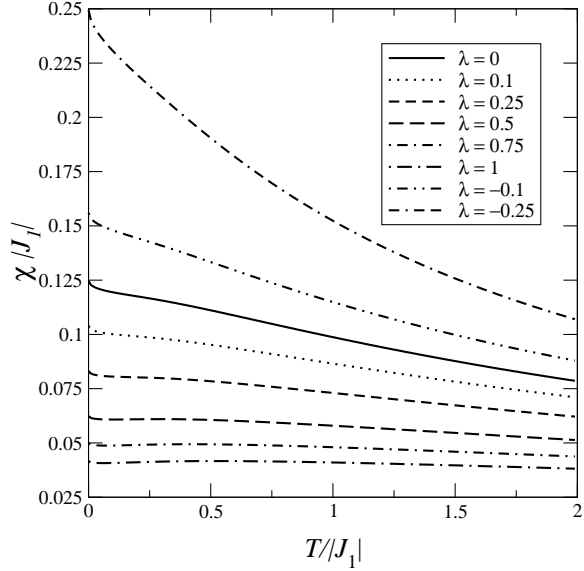} \par}

\caption{\label{fig.suscep.nnn}Effect of NNN interactions on the susceptibility
of the pyrochlore lattice.}
\end{figure}

Substituting back into the expression of the 1-spin cluster order
parameter we find the expression for the susceptibility\begin{equation}
\chi _{p}^{gcc}=\frac{1}{3\left| J_{1}\right| \widetilde{T}}\frac{1-(p-1)A_{p}(\widetilde{T})}{1+(1+z'\, \lambda )(p-1)A_{p}(\widetilde{T})}
\end{equation}
 or, in terms of the \( \varepsilon _{p} \) function defined above\begin{equation}
\chi _{p}^{gcc}=\frac{1}{3\left| J_{1}\right| \widetilde{T}}\frac{1+\varepsilon _{p}(\widetilde{T})}{1-(1+z'\, \lambda )\varepsilon _{p}(\widetilde{T})}.
\end{equation}
 This susceptibility has been depicted in Fig.~\ref{fig.suscep.nnn}
for different values of \( \lambda  \) and antiferromagnetic NN interactions.

\begin{figure}[htb!]
{\centering \subfigure[$J_1 < 0$]{\includegraphics{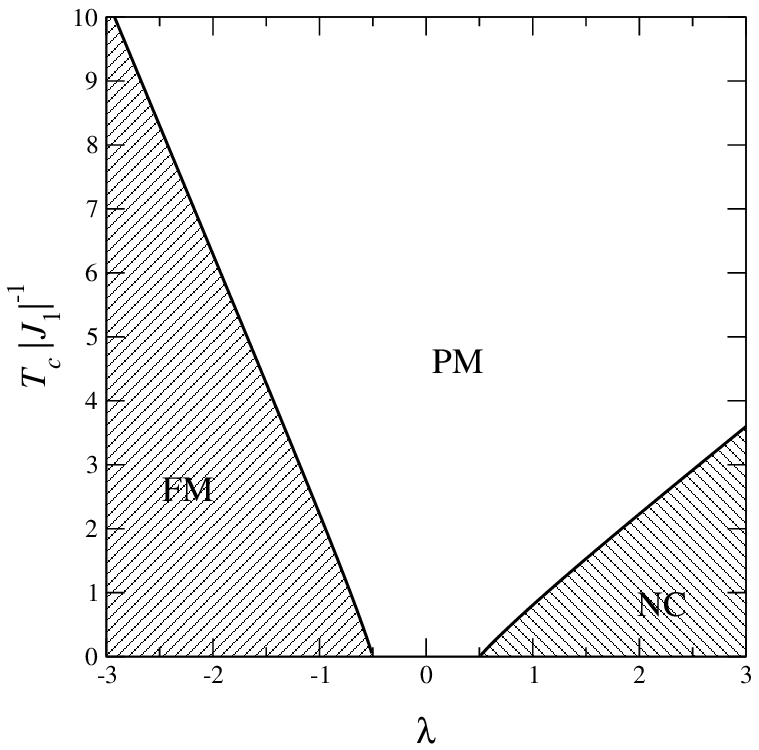}} \subfigure[$J_1 > 0$]{\includegraphics{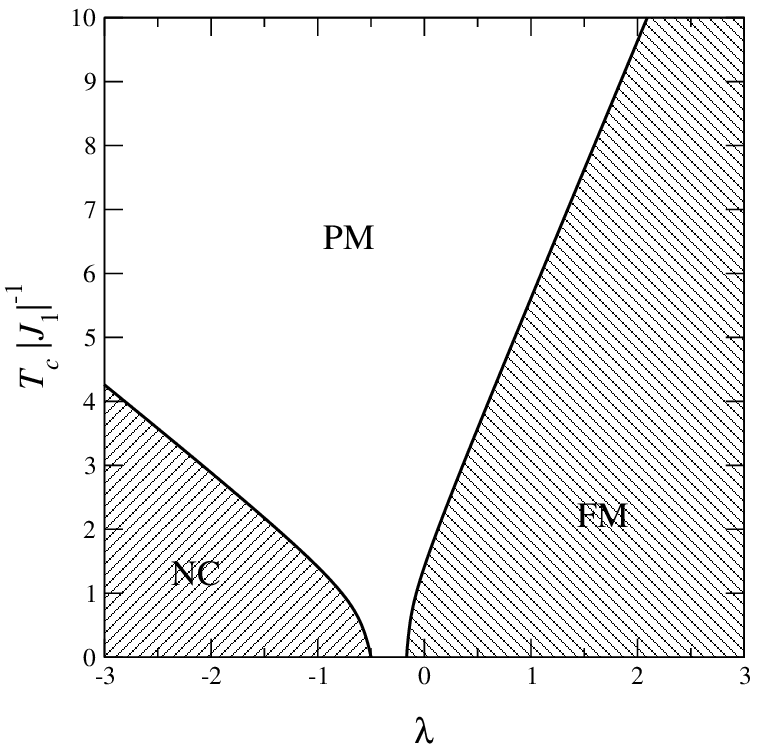}} \par}

\caption{\label{fig.phase.diagram.NNN}Phase diagram for the pyrochlore lattice
with NNN interactions. The curves are obtained for a fixed value of
\protect\( J_{1}\protect \).}
\end{figure}

In zero applied field, the system \eqref{nnn.system.eq} has two non-trivial
solutions, similar to the ones found for NN interactions only. The
ferromagnetic solution occurs if \begin{equation}
\varepsilon _{p}(\widetilde{T})=\frac{1}{(1+z'\, \lambda )},
\end{equation}
 and the non-collinear solution occurs if \begin{equation}
\varepsilon _{p}(\widetilde{T})=-\frac{(p-1)}{1+z'\, \lambda }.
\end{equation}
 These conditions have been studied and the main results are summarized
in Fig.~\ref{fig.phase.diagram.NNN}. As we can see from those figures,
NNN interactions can stabilize different ordered states depending
on the value of \( \lambda  \) and the sign of the NN interaction.
For antiferromagnetic NN interactions and \( \left| J_{2}\right| <0.5\left| J_{1}\right|  \),
the system remains paramagnetic down to 0 K. For antiferromagnetic
NN and NNN interactions, a NC ordered estate is selected at finite
temperature for \( \lambda >\frac{1}{2} \). However, for antiferromagnetic
NN interactions, and ferromagnetic NNN interactions, a ferromagnetic
ordered state is selected for \( \lambda <-\frac{1}{2} \). On the
other hand, the case of ferromagnetic NN interactions is slightly
more complex. For ferromagnetic NNN interactions, a ferromagnetic
ordered state is selected for \( -\frac{1}{6}<\lambda  \). In the
range \( -\frac{1}{2}<\lambda <-\frac{1}{6} \), the system remains
paramagnetic down to 0 K and, in the range \( \lambda <-\frac{1}{2} \),
a NC ordered state is selected.

In any case, from the analysis of this section, it seems very unlikely
that the non collinear ordered states experimentally found in some
pyrochlore systems are due to NNN interactions\cite{RAJU1999, GARCIA-ADEVA2000},
as in real pyrochlore systems \( \left| J_{2}\right| \sim 0.1\left| J_{1}\right|  \),
a value well inside the region of the phase diagram in which the system
is found to remain paramagnetic down to 0 K.

It is also important to stress that the present treatment of NNN interactions
is only approximate. In fact, in order to treat this problem consistently,
we should include new sublattices such as spins in one sublattice
do not interact with spins on the same sublattice. In doing so, it
can happen that new kinds of ordered states arise\cite{Reimers1991, RAJU1999}.
However, our main intention here is to check if the NC state given
by \eqref{condition.nc} is selected by NNN interactions. The general
problem with more sublattices will be studied elsewhere.

\section{Effects of site dilution by non-magnetic impurities}

\label{sec.dilution}

In order to study the effect of site dilution on the susceptibility
and on the conditions for the appearance of an ordered state, we need
to average both the SBF and the order parameters over the distribution
of non-magnetic impurities. For simplicity, we will assume that the
distribution of non-magnetic impurities is purely random. Under this
condition, the number of units with \( q \) spins, for a lattice
which, in the absence of dilution, is formed by units with \( p \)
spins, for a concentration of non-magnetic impurities \( x \), is
given by\cite{Moessner99}\begin{equation}
P_{q}^{p}(x)=\binom{p}{q}(1-x)^{q}x^{p-q}.
\end{equation}
 However, this distribution is normalized with respect to units, whereas
we need a distribution function normalized with respect to a single
spin. The adequate distribution function, for the SBF, is then\begin{equation}
\label{distri.func.sbf}
\mathcal{P}_{p}^{p}(x)=\frac{q}{p(1-x)}P_{q}^{p}(x),
\end{equation}
 and verifies the important relations we will use below\begin{equation}
\sum _{q=1}^{p}\mathcal{P}_{q}^{p}(x)=1,
\end{equation}
 \begin{equation}
\sum _{q=1}^{p}(q-1)\mathcal{P}_{q}^{p}(x)=(p-1)(1-x).
\end{equation}
 The corresponding distribution function for averaging over the order
parameter, however, has a different normalization. Indeed, if we average
the SBF for the \( 1- \)spin cluster with ferromagnetic interactions
with respect to \eqref{distri.func.sbf} we have\begin{equation}
\sum _{q=1}^{p}\mathcal{P}_{q}^{p}(x)\, \left\langle \vec{\xi }_{1\alpha }\right\rangle =\frac{\vec{h}_{0}}{\left| J\right| \widetilde{T}}+\frac{2(p-1)(1-x)}{\left| J\right| \widetilde{T}}\vec{h}',
\end{equation}
in agreement with the intuitive idea that, upon dilution, the applied
field is unaffected, and the EF changes according with the reduction
of the average coordination number by a factor \( (1-x) \), indicating
that the distribution function for the SBF is properly normalized.
If we now average the order parameter with respect to the same distribution
function, we find\begin{equation}
\sum _{q=1}^{p}\mathcal{P}_{q}^{p}(x)\, \vec{m}_{q\alpha }=\frac{\vec{h}_{0}}{3\left| J\right| \widetilde{T}}+\frac{2(p-1)(1-x)}{3\left| J\right| \widetilde{T}}\vec{h}',
\end{equation}
 and this result is incorrect, as it does not take into account the
reduction in the total number of magnetic ions in the lattice upon
dilution. Therefore, the correctly normalized distribution function
we have to use in order to average over the order parameter is given
by\begin{equation}
\label{distri.func.order.param}
\mathcal{Q}_{q}^{p}(x)=(1-x)\mathcal{P}_{q}^{p}(x).
\end{equation}

In the following, we will use the following notation to denote averages
with respect to dilution\begin{equation}
\left[ f_{q}\right] _{p}=\sum _{q=1}^{p}\mathcal{P}_{q}^{p}(x)\, f_{q},
\end{equation}
 where \( f_{q} \) is any quantity, and it will be understood that
we are using the distribution \eqref{distri.func.sbf} when averaging
over SBF and the distribution \eqref{distri.func.order.param} when
averaging over the order parameter.

Furthermore, for the sake of clarity, we will consider the antiferromagnetic
case with no applied field separately from the paramagnetic and ferromagnetic
with no applied field cases. In the paramagnetic case and ferromagnetic
case in the absence of an applied field, we have seen that the system
can be described by a single effective field, \( \vec{h}' \). The
average with respect to dilution of the SBF for the 1-spin cluster,
as we have seen above, is given by\begin{equation}
\left[ \left\langle \vec{\xi }_{1\alpha }\right\rangle \right] _{p}=\frac{2(p-1)(1-x)}{\left| J\right| \widetilde{T}}\vec{h}'+\frac{\vec{h}_{0}}{\left| J\right| \widetilde{T}},
\end{equation}
 where we have only considered the contribution of NN to the SBF.
Now, the averaged 1-spin order parameter is given by\begin{equation}
\left[ \vec{m}_{1\alpha }\right] _{p}=\frac{1-x}{3\left| J\right| \widetilde{T}}\left[ \vec{h}_{0}+2(p-1)(1-x)\, \vec{h}'\right] .
\end{equation}
 Analogously, for the \( p \)-spin cluster we have\begin{equation}
\left[ \left\langle \vec{\xi }_{q\alpha }\right\rangle \right] _{p}=\frac{(p-1)(1-x)}{\left| J\right| \widetilde{T}}\vec{h}'+\frac{\vec{h}_{0}}{\left| J\right| \widetilde{T}},
\end{equation}
 \begin{equation}
\sum _{\beta \neq \alpha }\left[ \left\langle \vec{\xi }_{q\beta }\right\rangle \right] _{p}=\frac{q-1}{\left| J\right| \widetilde{T}}\left( (p-1)(1-x)\vec{h}'+\vec{h}_{0}\right) ,
\end{equation}
 and the corresponding averaged order parameter is given by\begin{equation}
\left[ \vec{m}_{q\alpha }\right] _{p}=\frac{1-x}{3\left| J\right| \widetilde{T}}\left( 1-\bar{A}_{p}(\widetilde{T})\right) \left( \vec{h}_{0}+(p-1)(1-x)\vec{h}'\right) ,
\end{equation}
 where \begin{equation}
\bar{A}_{p}(\widetilde{T})=\sum _{q=1}^{p}\mathcal{P}_{q}^{p}(x)\, (q-1)\, A_{q}(\widetilde{T}).
\end{equation}
 Therefore the averaged self-consistency condition \begin{equation}
\left[ \vec{m}_{1\alpha }\right] _{p}=\left[ \vec{m}_{q\alpha }\right] _{p},
\end{equation}
leads us to the following equation for the effective field \( \vec{h}' \)\begin{equation}
\label{self.consistency.dilution}
\vec{h}_{0}+2(p-1)(1-x)\vec{h}'=(1-\bar{A}_{p})\left( \vec{h}_{0}+(p-1)(1-x)\vec{h}'\right) ,
\end{equation}
 that has the solution\begin{equation}
\vec{h}'=-\frac{\bar{A}_{p}}{(p-1)(1-x)(1+\bar{A}_{p})}\vec{h}_{0},
\end{equation}
 and the corresponding susceptibility is found to be\begin{equation}
\left[ \chi \right] _{p}=\frac{1-x}{3\left| J\right| \widetilde{T}}\frac{1-\bar{A}_{p}}{1+\bar{A}_{p}},
\end{equation}
 which is of the same form as \eqref{suscep.nnn}, except for that
\( (p-1)A_{p} \) is replaced by the corresponding average over dilution.
Interestingly, it is easily shown that this average can be expressed
in terms of an average over the \( \varepsilon _{p} \) function introduced
before. Indeed, if we define\begin{equation}
\bar{\varepsilon }_{p}=\frac{2\widetilde{T}^{2}}{p(1-x)}\frac{\partial }{\partial \widetilde{T}}\ln \bar{z}^{0}_{p}(\widetilde{T},x)-1,
\end{equation}
 where \( \bar{z}_{p}^{0} \) is the average of \( z_{p}^{0} \) with
respect to dilution, given by\begin{equation}
\bar{z}_{p}^{0}=\prod _{q=1}^{p}z_{p}^{0}(\widetilde{T})^{P_{q}^{p}(x)},
\end{equation}
 we have the identity\begin{equation}
\label{rel1}
\bar{\varepsilon }_{p}=-\bar{A}_{p},
\end{equation}
 and we can put the averaged susceptibility in the form\begin{equation}
\left[ \chi \right] _{p}=\frac{1-x}{3\left| J\right| \widetilde{T}}\frac{1+\bar{\varepsilon }_{p}}{1-\bar{\varepsilon }_{p}}.
\end{equation}
 Incidentally, this is precisely the susceptibility calculated by
the present authors in Ref.~\onlinecite{GARCIA-ADEVA2001d}, so we
recover all the results presented there for the paramagnetic case.
\begin{figure}[htb!]
{\centering \subfigure[Ferromagnetic phase diagram]{\includegraphics{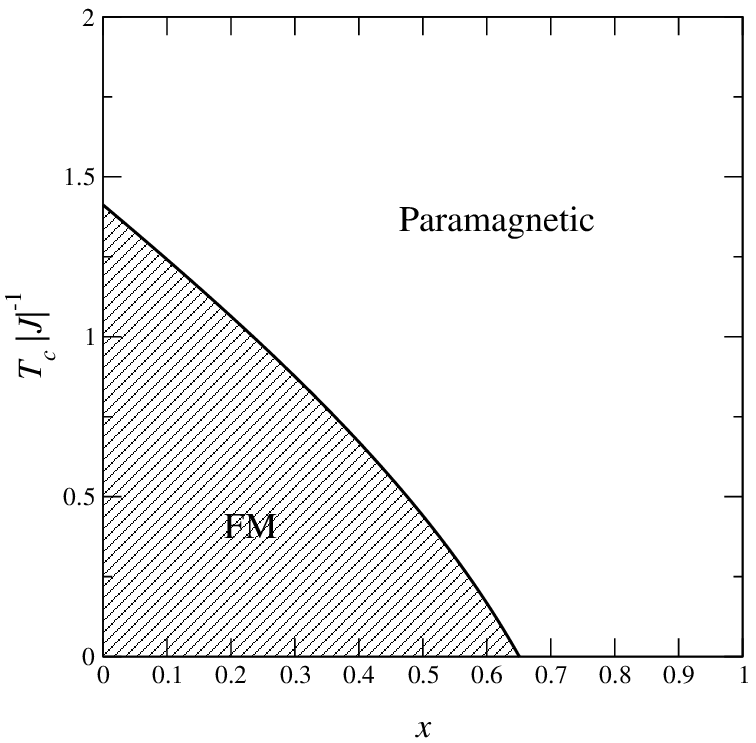}} \subfigure[\label{fig.cond.nc.dilution}Condition for non-collinear order]{\includegraphics{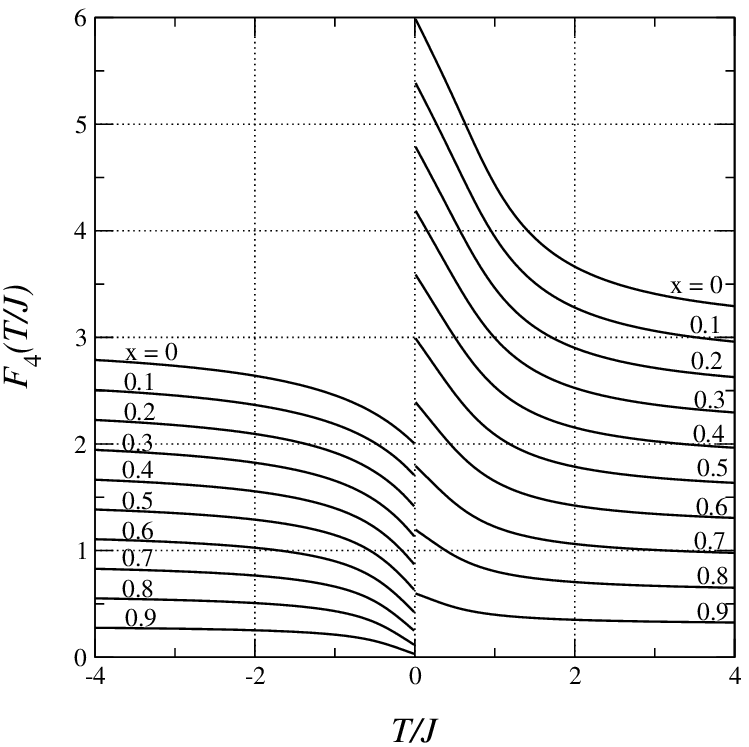}} \par}

\caption{\label{fig.dilution}(a) Phase diagram for the ferromagnetic pyrochlore
with site dilution. (b) Study of the condition for the existence of
non-collinear order. \protect\( F_{4}\protect \) stands for the function
\protect\( F_{p}=\bar{\varepsilon }_{p}-(p-1)(1-x)\protect \) for
\protect\( p=4\protect \) (see text).}
\end{figure}

In the absence of applied field, it can be easily shown from eq. \eqref{self.consistency.dilution}
that the condition for the formation of a ferromagnetic ordered state
is given by\begin{equation}
\bar{\varepsilon }_{p}=1.
\end{equation}
 To check the consistency of the results, we have studied this condition
for different amounts of dilution for the ferromagnetic pyrochlore,
and the dependence of the transition temperature with \( x \) is
depicted in Fig.~\ref{fig.dilution}. We can see that there is a critical
value \( x_{c} \) above which there is no transition. This value
is in reasonable agreement with the value of the percolation threshold
for the pyrochlore lattice, \( x_{c}\simeq 0.6 \), whereas our model
predicts a value \( x_{c}\simeq 0.66 \). This result is reasonable,
as the transition temperatures predicted by our model are slightly
larger than the real ones, due to neglecting long wavelength fluctuations.

Let us now consider the antiferromagnetic case with no applied field,
and let us calculate the condition for the existence of non-collinear
order upon dilution. In this case, we know from the analysis in the
previous sections that the effective fields \( \vec{h}'_{\alpha } \)
verify the relation\begin{equation}
\sum _{\beta \neq \alpha }\vec{h}'_{\beta }=-\vec{h}'_{\alpha }
\end{equation}
 at the transition point, if any. Therefore, we have the following
expressions for the dilution averaged SBF in this case\begin{equation}
\left[ \left\langle \vec{\xi }_{1\alpha }\right\rangle \right] _{p}=2\, \left[ \left\langle \vec{\xi }_{q\alpha }\right\rangle \right] _{p}=-\frac{2\vec{h}'_{\alpha }}{\left| J\right| \widetilde{T}},
\end{equation}
 and\begin{equation}
\sum _{\beta \neq \alpha }\left[ \left\langle \vec{\xi }_{q\beta }\right\rangle \right] _{p}=\frac{\vec{h}'_{\alpha }}{\left| J\right| \widetilde{T}}.
\end{equation}
 The order parameter for a unit with \( q<p \) spins is then given
by\begin{equation}
\vec{m}_{q\alpha }=\frac{1}{3\left| J\right| \widetilde{T}}\left[ 1+A_{q}\right] \vec{h}'_{\alpha }.
\end{equation}
 If we now naively average the self consistency condition as above,
we would not be able to obtain the condition for the transition to
a non-collinear state in terms of the \( \bar{A}_{p} \) or \( \bar{\varepsilon }_{p} \)
functions introduced before, as a \( (q-1) \) prefactor is lacking
in the expression for \( \vec{m}_{q\alpha } \). Thus, we construct
the following averages instead\begin{equation}
\left[ (q-1)\, \vec{m}_{1\alpha }\right] _{p}=-\frac{2(p-1)(1-x)}{3\left| J\right| \widetilde{T}}\vec{h}'_{\alpha },
\end{equation}
 and\begin{equation}
\left[ (q-1\, )\vec{m}_{q\alpha }\right] _{p}=-\frac{1-x}{3\left| J\right| \widetilde{T}}\left( (p-1)(1-x)+\bar{A}_{p}\, \right) \, \vec{h}'_{\alpha }.
\end{equation}
 Equating both quantities, we easily obtain the condition for the
existence of non-collinear order upon site dilution\begin{equation}
\bar{\varepsilon }_{p}=-(p-1)(1-x),
\end{equation}
 where we have made use of relation \eqref{rel1}.

As can be seen from Fig.~\ref{fig.cond.nc.dilution}, the function
\( F_{p}=\bar{\varepsilon }_{p}-(p-1)(1-x) \) does not change its
sign for any value of \( x \), indicating that a non-collinear ordered
state is not selected by site dilution, in contrast with the qualitative
picture suggested in other works.

\section{Combined effect of NNN interactions and site dilution}

\label{sec.nnn.dilution}

Of course, in real systems, the effects analyzed above are always
present simultaneously. Therefore, it is important to study how the
conclusions reached in the previous sections are modified when we
simultaneously consider both NNN interactions and site dilution. We
will only present the main results of the analysis, as the derivation
of the final expressions follows very closely the previous derivations.

The averaged susceptibility is obtained to be\begin{equation}
\left[ \chi \right] _{p}=\frac{1-x}{3\left| J_{1}\right| \widetilde{T}}\, \frac{1+\bar{\varepsilon }_{p}}{1-(1+z'\, \lambda )\bar{\varepsilon }_{p}}.
\end{equation}

\begin{figure}[htb!]
{\centering \subfigure[$J_1<0$]{\includegraphics{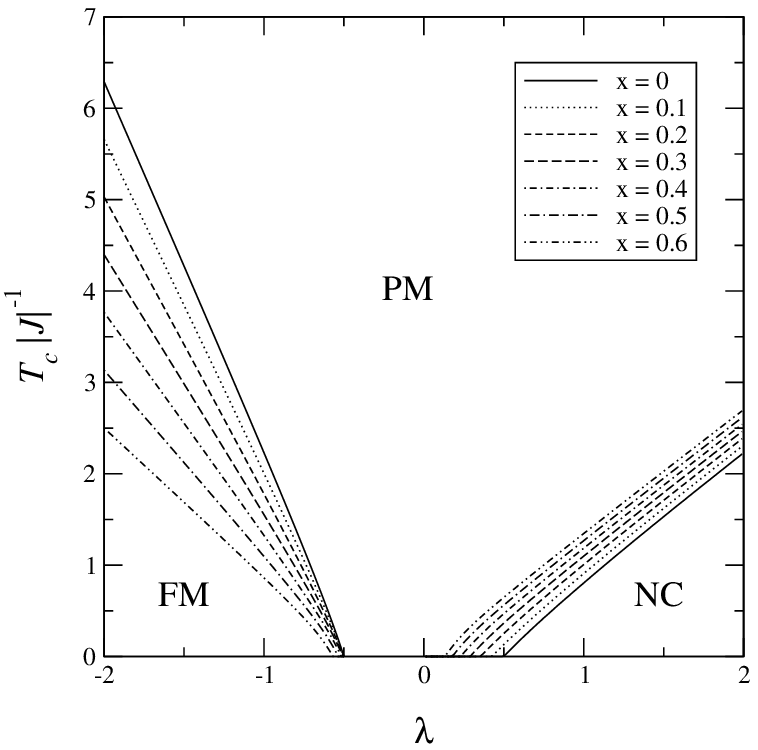}} \subfigure[$J_1>0$]{\includegraphics{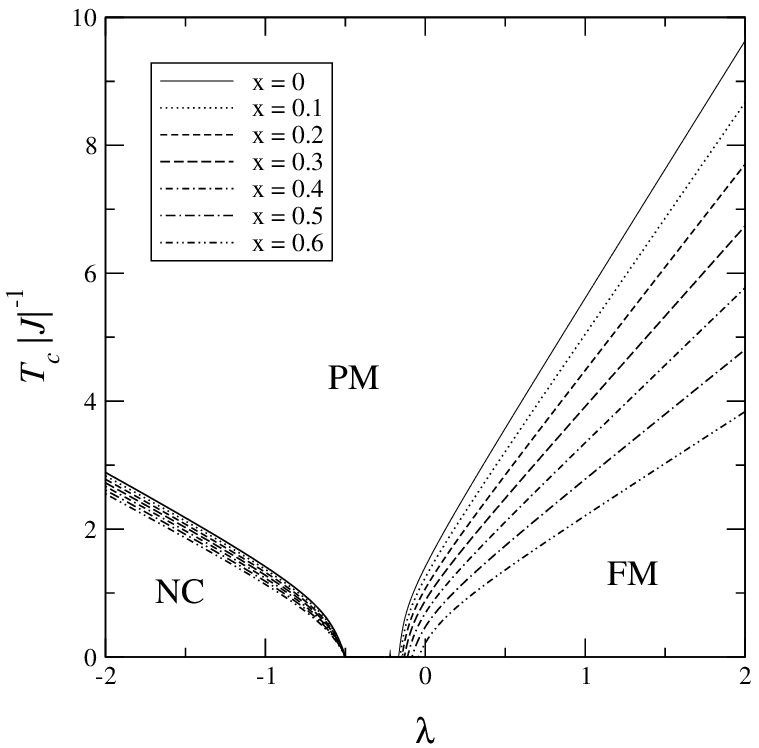}} \par}

\caption{\label{fig.phase.diagram.nnn.dilution}Phase diagrams with both NNN
interactions and site dilution.}
\end{figure}

\begin{figure}[htb!]
{\centering \includegraphics{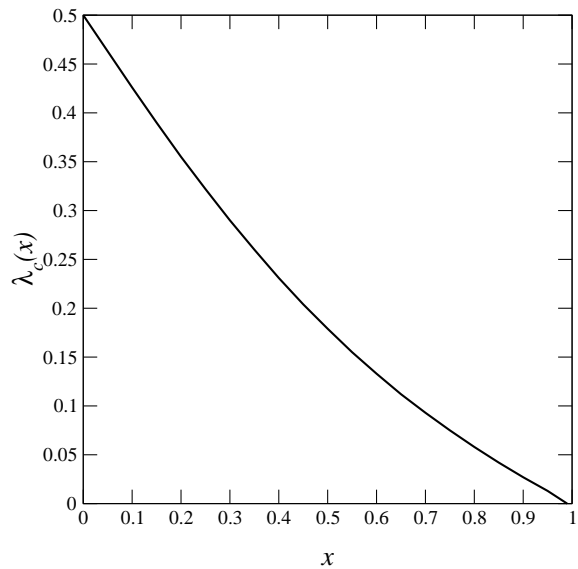} \par}

\caption{\label{fig.lambdac}Minimum value of \protect\( \lambda \protect \)
for which a NC ordered state is formed for different concentrations
of non-magnetic impurities.}
\end{figure}

The condition for the existence of a ferromagnetic ordered state is
given by \( \bar{\varepsilon }_{p}=\frac{1}{1+z'\, \lambda }, \)
whereas the corresponding one for a non-collinear ordered state is
given by \( \bar{\varepsilon }_{p}=-\frac{(p-1)(1-x)}{1+z'\, \lambda }. \)
The phase diagrams, for different concentrations of non-magnetic impurities
are depicted in Fig.~\ref{fig.phase.diagram.nnn.dilution}. Also,
in Fig.~\ref{fig.lambdac} we have depicted the values of \( \lambda _{c}(x) \),
that is, the value of \( \frac{J_{2}}{J_{1}} \) above which a non-collinear
ordered state is formed at a finite temperature, for a given value
of \( x \). We have only depicted this quantity for antiferromagnetic
NN interactions, as the corresponding \( \lambda _{c} \) for ferromagnetic
NN interactions does not change upon dilution.

\section{Summary and Conclusions}

\label{sec.conclusions}

In this work we have presented a microscopic formulation of the Generalized
Constant Coupling method for geometrically frustrated magnets. By
introducing different Bravais sublattices characterized by different
order parameters we are able to describe more general kinds of long
range ordered states, including non-collinear ones. These order parameters
are evaluated by making use of the Suzuki-Callen identity for finite
size clusters, in terms of symmetry breaking fields arising from interactions
with units surrounding the considered one. The symmetry breaking fields
are easily related with the phenomenological effective fields introduced
in the original formulation of the generalized constant coupling method,
in terms of averages over the neighboring spin variables outside the
unit, and they are fixed by a self-consistency condition. Though we
have not stressed this point, this self-consistency condition is,
in fact, a particular case of the more general scaling hypothesis
between the order parameters for finite clusters of different sizes\cite{GARCIA-ADEVA2001c}.

When a magnetic field is applied to the system, this method allows
us to easily compute the susceptibility, and we recover previous results
obtained in the phenomenological formulation of the GCC method, which
were shown to be in excellent agreement with MC results for both the
kagome and pyrochlore lattice. However, the present formulation, allows
us to also study the critical behavior of the system when no applied
magnetic field is present. The possible ordered configurations for
these systems naturally arise from the equations that fix the order
parameters. In completely agreement with the accepted idea, for ferromagnetic
interactions a ferromagnetic configuration in which all the spins
point along the same direction is found. For antiferromagnetic interactions,
a non-collinear state in which the total magnetization of each cluster
is identically zero is found. Also, the conditions for the formation
of such ordered states arise from the calculation. It is found that,
for nearest neighbor antiferromagnetic interactions, frustration inhibits
the formation of a long range ordered state, and the system remains
paramagnetic down to 0 K. We have briefly compared the results for
these frustrated geometries with the results for a simple cubic lattice
obtained with this same method, and we found that the generalized
constant coupling method correctly distinguishes between different
topologies of the lattice, even in case where the number of nearest
neighbors is the same, as it happens with the pyrochlore and cubic
lattices. Another advantage of the present formulation is that it
is constructed in the real space, which makes it very easy to include
additional perturbations to the system, by modifying the form of the
symmetry breaking fields. Particularly, we have studied the effect
of next nearest neighbor interactions and site dilution by non-magnetic
impurities for the pyrochlore lattice. In the former case, the phase
diagram has been calculated. It is found that, for certain values
of the ratio \( \lambda =\frac{J_{2}}{J_{1}} \), a non-collinear
or ferromagnetic ordered state can be stabilized. However, the minimum
value of \( \lambda  \) above which a non-collinear configuration
is estable seems to be too large to explain the formation of long
range order found in some systems, notably in Gd\( _{2} \)Ti\( _{2} \)O\( _{7} \)\cite{RAJU1999}.
Moreover, it seems clear from neutron scattering that the ordered
configuration in this case is not compatible with the condition that
the total magnetization of the tetrahedral cluster is zero\cite{CHAMPION5043}.
However, the main perturbation in this system seems to come from long
range dipolar interactions, which are highly anysotropic. Further
theoretical work in this direction, in the present framework, is still
necessary.

Regarding the effects of site dilution, it has been argued that this
kind of effect, always present in real materials, could break the
non-trivial degeneracy of the ground state and cause a transition
to some kind of long range ordered state at finite temperature. However,
in our calculation, we find that this is not case: Site dilution by
itself does not induce a transition, for any amount of dilution. It
is necessary to include simultaneously both dilution and next nearest
neighbor interactions.

A word of caution must be said here, however. In the last part of
this work, we have focused our attention on studying if a non-collinear
ordered state given by the rule that the total magnetization of the
tetrahedral unit is zero is estabilized by any of these perturbations.
However, these are not the only possible ordered configurations when
we include these effects. For example, Reimers and coworkers\cite{Reimers1991}
showed that, in the framework of mean field theory, next nearest neighbor
interactions can induce both non-collinear ordered states as the ones
studied in this work, or ordered states characterized by an inconmensurate
wave vector. It is not easy to describe these inconmesurate states
with a real space method, and it could be the case that site dilution
would lead to such an inconmesurate state. Therefore, more theoretical
work in that direction should be necesary before ruling out the possibility
of a long range ordered state induced by small amounts of site dilution
or, what would be even more interesting, the possibility of a transition
to some kind of spin glass state induced by dilution\cite{VILLAIN1979}.

In conclusion, the formulation of the generalized constant coupling
method presented in this work provides us a conceptual framework in
which both thermodynamic quantities and the critical behavior of geometrically
frustrated magnets can be properly described. The method can be easily
generalized to deal with further perturbations always present in real
systems. We have presented the classical limit of the method. However,
in order to compare the calculated quantities with experimental data
for these systems, the corresponding quantum generalized constant
coupling method should be used. This issue, the occurrence of inconmensurate
states due to dilution, and the effect of the inclusion of long range
dipolar interactions in the Hamiltonian are open issues which deserve
further theoretical work.

\appendix

\section{Evaluation of the  \emph{p}-spin cluster partition function.}
\label{appendix1}

In this appendix we present the main step for the evaluation of the
integral \eqref{partition.cluster}\begin{equation}
Z_{p}(K)=\int \prod _{\alpha }d\Omega _{\alpha }e^{\mathcal{H}_{p}},
\end{equation}
 which can be put in the more convenient form\begin{equation}
Z_{p}(K)=\int d^{3}\vec{S}\int \prod _{\alpha }d^{3}\vec{s}_{1\alpha }\, \delta \left( \vec{S}-\sum _{\alpha }\vec{s}_{1\alpha }\right) \, \exp \left( K\sum _{\alpha \neq \beta }\vec{s}_{1\alpha }\cdot \vec{s}_{1\beta }+\sum _{\alpha }\vec{s}_{1\alpha }\cdot \vec{\xi }_{p\alpha }\right) .
\end{equation}
 By using the Fourier transform representation of the Dirac-\( \delta  \)\begin{equation}
\delta (\vec{x})=\frac{1}{(2\pi )^{3}}\int d^{3}\vec{q}\, e^{i\vec{q}\cdot \vec{x}},
\end{equation}
 we can further put\begin{equation}
Z_{p}(K)=\frac{1}{(2\pi )^{3}}\int d^{3}\vec{q}\int d^{3}\vec{S}\, e^{\frac{K}{2}S^{2}}\, e^{i\vec{q}\cdot \vec{S}}\prod _{\alpha }\int d^{3}\vec{s}_{\alpha }\, e^{\vec{a}_{\alpha }\cdot \vec{s}_{\alpha }},
\end{equation}
 where we have introduced the vector \( \vec{a}=\vec{\xi }_{p\alpha }-i\vec{q} \),
and made use of the fact\begin{equation}
\frac{1}{2}\sum _{\alpha \neq \beta }\vec{s}_{1\alpha }\cdot \vec{s}_{1\beta }=\frac{S^{2}}{2},
\end{equation}
 with \( \vec{S} \) the total spin of the unit, up to an unimportant
global constant in the Hamiltonian. The integral over \( \vec{s}_{\alpha } \)
can be done as follows. Taking into account\begin{equation}
\vec{a}\cdot \vec{s}=a_{x}\sin \theta \cos \varphi +a_{y}\sin \theta \sin \varphi +a_{z}\cos \theta ,
\end{equation}
 where we have assumed that the length of \( \vec{s}_{\alpha } \)
is 1 (this can be always done by redefining \( K\to Ks_{0}^{2} \)
with \( s_{0} \) the length of the spin), the integral over \( \vec{s}_{\alpha } \)
can be put as\\
\begin{multline} \int _{0}^{\pi }d\theta \int _{0}^{2\pi }d\varphi \, \sin \theta \, e^{a_{z}\cos \theta +a_{y}\sin \theta \sin \varphi +a_{z}\sin \theta \cos \varphi }=2\pi \int _{0}^{\pi }d\theta \, \sin \theta \, e^{a_{z}\cos \theta }\, I_{0}\left( \sqrt{a_{x}^{2}+a_{y}^{2}}\sin \theta \right)\\ 
=2\pi \int _{-1}^{-1}dx\, e^{-ax}\, I_{0}\left( b\sqrt{1-x^{2}}\right)=4\pi\frac{\sinh\sqrt{a^2+b^2}}{\sqrt{a^2+b^2}}= 4\pi\frac{\sinh\sqrt{a_x^2+a_y^2+a_z^2}}{\sqrt{a_x^2+a_y^2+a_z^2}}=4\pi\frac{\sin a_\alpha}{a_\alpha},
\end{multline}
where \( \vec{a}_{\alpha }=\vec{q}+i\, \vec{\xi }_{p\alpha } \) (we
have factored out a (-1) in the definition of \( \vec{a}_{\alpha } \)).

Next, we can do the integral over \( \vec{S} \) \begin{equation}
\int d^{3}\vec{S}\, e^{\frac{K}{2}S^{2}}\, e^{i\vec{q}\cdot \vec{S}}=4\pi \frac{\sqrt{\pi /2}}{(-K)^{3/2}}e^{-q^{2}/2(-K)},
\end{equation}
 which only converges for \( K<0 \), and we are left with an integral
over \( \vec{q} \)\begin{equation}
\label{zp.first}
Z_{p}(K)\sim \int d^{3}\vec{q}\, e^{\frac{q^{2}}{2K}}\, \prod _{\alpha }\frac{\sin \sqrt{q^{2}+2i\vec{q}\cdot \vec{\xi }_{p\alpha }-\xi _{p\alpha }^{2}}}{\sqrt{q^{2}+2i\vec{q}\cdot \vec{\xi }_{p\alpha }-\xi _{p\alpha }^{2}}},
\end{equation}
 which cannot be calculated in a closed form. An interesting point
here is that, if we take \( \vec{\xi }_{p\alpha }=\vec{0} \), we
recover the partition function calculated in the original GCC model.
The only thing we can do at this point is to expand the partition
function in powers of the SBF, up to second order, as the rest of
terms will be assumed to be small near the critical point once we
take the thermal average. With a little algebra it is easy to show
that\begin{equation}
\frac{\sin \sqrt{q^{2}+2i\vec{q}\cdot \vec{\xi }_{p\alpha }-\xi _{p\alpha }^{2}}}{\sqrt{q^{2}+2i\vec{q}\cdot \vec{\xi }_{p\alpha }-\xi _{p\alpha }^{2}}}\simeq f_{0}(q)+i\, f_{1}(q)\, \xi _{p\alpha }\cos \theta _{\alpha }+f_{2}(q)\, \xi _{p\alpha }^{2}+f_{3}(q)\, \xi _{p\alpha }^{2}\cos ^{2}\theta _{\alpha },
\end{equation}
 where \( \theta _{\alpha } \) is the angle defined by \( \vec{\xi }_{p\alpha } \)
and \( \vec{q} \), and\begin{equation}
f_{0}(q)=\frac{\sin q}{q}
\end{equation}
 \begin{equation}
f_{1}(q)=\frac{\cos q}{q}-\frac{\sin q}{q^{2}}
\end{equation}
 \begin{equation}
f_{2}(q)=\frac{\sin q}{2q^{3}}-\frac{\cos q}{2q^{2}}
\end{equation}
 \begin{equation}
f_{3}(q)=\frac{3\cos q}{2q^{2}}-\frac{3\sin q}{2q^{3}}+\frac{\sin q}{2q}.
\end{equation}

By making use of this result, we can put 
\begin{multline} 
\label{series1} 
\prod _{\alpha }\frac{\sin \sqrt{q^{2}+2i\vec{q}\cdot \vec{\xi }_{p\alpha }-\xi _{p\alpha }^{2}}}{\sqrt{q^{2}+2i\vec{q}\cdot \vec{\xi }_{p\alpha }-\xi _{p\alpha }^{2}}}\simeq f_{0}^{p}+f_{0}^{p-1}f_{2}\sum _{\alpha }\xi _{p\alpha }^{2}+f_{0}^{p-1}f_{3}\sum _{\alpha }\cos ^{2}\theta _{\alpha }\xi _{p\alpha }^{2}\\ 
-\frac{f_{0}^{p-2}f_{1}^{2}}{2}\sum _{\alpha \neq \beta }\cos \theta _{\alpha }\cos \theta _{\beta }\xi _{p\alpha }\xi _{p\beta }+i\, f_{0}^{p-1}f_{1}\sum _{\alpha }\cos \theta _{\alpha }\xi _{p\alpha }. 
\end{multline}\\
The expressions of the form \( \xi _{p\alpha }\cos \theta _{\alpha } \)
are the projection of \( \vec{\xi }_{p\alpha } \) over \( \vec{q} \),
and can be put in the more convenient form\begin{equation}
\xi _{p\alpha }\cos \theta _{\alpha }=\xi _{p\alpha }^{x}\sin \theta \cos \varphi +\xi _{p\alpha }^{y}\sin \theta \sin \varphi +\xi _{p\alpha }^{z}\cos \theta .
\end{equation}
 If we consider the imaginary part of the previous series, and integrate
over the angular part of \( \vec{q} \) in \eqref{zp.first}, we have
integrals of the form \begin{equation}
\int _{0}^{\pi }d\theta \int _{0}^{2\pi }d\varphi \, \sin \theta \, \xi _{p\alpha }\cos \theta _{\alpha },
\end{equation}
 which are easily shown to be zero. 

We also have integrals of the form\begin{equation}
\int _{0}^{\pi }d\theta \int _{0}^{2\pi }d\varphi \, \sin \theta \, (\xi _{p\alpha }\cos \theta _{\alpha })^{2}=\frac{4\pi }{3}\left[ \left( \xi _{p\alpha }^{x}\right) ^{2}+\left( \xi _{p\alpha }^{y}\right) ^{2}+\left( \xi _{p\alpha }^{z}\right) ^{2}\right] =\frac{4\pi }{3}\xi _{p\alpha }^{2}
\end{equation}
 and\begin{equation}
\int _{0}^{\pi }d\theta \int _{0}^{2\pi }d\varphi \, \sin \theta \, \xi _{p\alpha }\cos \theta _{\alpha }\, \xi _{p\beta }\cos \theta _{\beta }=\frac{4\pi }{3}\left[ \xi _{p\alpha }^{x}\xi _{p\beta }^{x}+\xi _{p\alpha }^{y}\xi _{p\beta }^{y}+\xi _{p\alpha }^{z}\xi _{p\beta }^{z}\right] =\frac{4\pi }{3}\vec{\xi }_{p\alpha }\cdot \vec{\xi }_{p\beta }.
\end{equation}

Substituting these results back in \eqref{series1} we are left with
the integral over \( q \)\begin{equation}
Z_{p}(K)\sim \int _{0}^{\infty }dq\, q^{2}\, e^{\frac{q^{2}}{2K}}\left( f_{0}^{p}+f_{0}^{p-1}f_{2}\sum _{\alpha }\xi _{p\alpha }^{2}+\frac{f_{0}^{p-1}f_{3}}{3}\sum _{\alpha }\xi _{p\alpha }^{2}-\frac{f_{0}^{p-2}f_{1}^{2}}{6}\sum _{\alpha \neq \beta }\vec{\xi }_{p\alpha }\cdot \vec{\xi }_{p\beta }\right) 
\end{equation}
 which, taking into account \begin{equation}
f_{2}+\frac{f_{3}}{3}=\frac{f_{0}}{6}
\end{equation}
 can be simplified to the form\begin{equation}
Z_{p}\sim \int _{0}^{\infty }dq\, q^{2}\, e^{\frac{q^{2}}{2K}}\, \left( f_{0}^{p}+\frac{f_{0}^{p}}{6}\sum _{\alpha }\xi _{p\alpha }^{2}-\frac{f_{0}^{p-2}f_{1}^{2}}{6}\sum _{\alpha \neq \beta }\vec{\xi }_{p\alpha }\cdot \vec{\xi }_{p\beta }\right) .
\end{equation}
 Each of the integrals can be put as a closed expressions, and are
given in the next section. For the moment, we will use the notation
\begin{equation}
z^{0}_{p}(K)=\int _{0}^{\infty }dq\, q^{2}\, e^{\frac{q^{2}}{2K}}f_{0}^{p}
\end{equation}
 and\begin{equation}
z_{p}^{1}(K)=\int _{0}^{\infty }dq\, q^{2}\, e^{\frac{q^{2}}{2K}}f_{0}^{p-2}f_{1}^{2}.
\end{equation}

Finally, up to second order in the SBF, we can put\begin{equation}
Z_{p}(K)\sim z_{p}^{0}(K)+\frac{z_{p}^{0}(K)}{6}\sum _{\alpha }\xi _{p\alpha }^{2}-\frac{z_{p}^{1}(K)}{6}\sum _{\alpha \neq \beta }\vec{\xi }_{p\alpha }\cdot \vec{\xi }_{p\beta }.
\end{equation}

\section{Expressions for \protect\( z_{p}^{1}(K)\protect \)
and \protect\( z_{p}^{0}(K)\protect \)}
\label{appendix2}

The \( z_{p}^{0}(K) \) function for different values of \( p \)
and antiferromagnetic interactions (\( K<0 \)) is given as follows\begin{equation}
z_{1}^{0}(K)=e^{\frac{K}{2}},
\end{equation}
 \begin{equation}
z_{2}^{0}(K)=\frac{e^{2K}-1}{K},
\end{equation}
 \begin{equation}
z_{3}^{0}(K)=\frac{3\erf \left( \sqrt{\frac{-K}{2}}\right) +\erf \left( 3\sqrt{\frac{-K}{2}}\right) }{(-K)^{5/2}},
\end{equation}
\begin{equation}
z_{4}^{0}(K)=\frac{2\erf \left( \sqrt{-2K}\right) -\erf \left( \sqrt{-8K}\right) }{(-K)^{3/2}}-\frac{3+e^{8K}-4e^{-2K}}{\sqrt{8\pi }\, K^{2}}.
\end{equation}
 In these expressions, \( \erf  \) represents the error function.
It is important to notice that, if we make the replacement \( K\to -\frac{1}{T} \),
we immediately recover the expressions of Ref.~\onlinecite{Moessner99}.
For ferromagnetic interactions (\( K>0 \)), the expressions are completely
similar to the ones above, with the substitutions \( \sqrt{-K}\to \sqrt{K} \)
and \( \erf \to \erfi  \), with \( \erfi (x)=\frac{\erf (i\, x)}{i} \)
the imaginary error function. To quote an example\begin{equation}
z_{4}^{0}(K)=\frac{2\erfi \left( \sqrt{2K}\right) -\erfi \left( \sqrt{8K}\right) }{K^{3/2}}-\frac{3+e^{8K}-4e^{-2K}}{\sqrt{8\pi }\, K^{2}}.
\end{equation}

The \( z_{p}^{1}(K) \) functions are obtained to be, for antiferromagnetic
interactions\begin{equation}
z_{2}^{1}(K)=\frac{e^{K}}{K^{2}}\left( \sinh K-K\cosh K\right) ,
\end{equation}
 \begin{equation}
z_{3}^{1}(K)=\frac{e^{9K}\left( e^{-4K}-1\right) }{K^{2}}-\sqrt{\frac{\pi }{2}}\frac{1+K}{(-K)^{3/2}}\left[ 3\erf \left( \sqrt{\frac{-K}{2}}\right) -\erf \left( 3\sqrt{\frac{-K}{2}}\right) \right] ,
\end{equation}
 \begin{equation}
z_{4}^{1}(K)=\sqrt{\frac{\pi }{2}}\frac{4K+3}{(-K)^{5/2}}\left[ 2\erf \left( \sqrt{-2K}\right) -\erf \left( \sqrt{-8K}\right) \right] -\frac{(1+K)\left( 3+e^{8K}-4e^{2K}\right) }{K^{3}}.
\end{equation}
 The corresponding expressions for the ferromagnetic case can be obtained
by using the prescription indicated above. It is important to notice
that \( z_{p}^{1}(K) \) is not defined for \( p=1 \) but it easy
to see that this function does not appear in any calculation, as it
corresponds to non-interacting spins.

\begin{acknowledgments}
AJGA wants to thank the Spanish Ministerio de Ciencia y Tecnologia
for financial support under the Subprograma General de Formacion de
Doctores y Tecnologos en el Extranjero.
\end{acknowledgments}

\end{document}